\documentclass{sig-alternate-05-2015}

\usepackage{etex}
\usepackage[
bookmarks=false
,bookmarksnumbered=true
,hypertexnames=false
,breaklinks=true
]{hyperref}
\hypersetup{
  pdfauthor={Dehghani, Azarbonyad, Kamps, Marx},
  pdftitle={On Horizontal and Vertical Separation in Hierarchical Text Classification},
  pdfsubject={ICTIR'16 submission},
  pdfkeywords={H.3 [Information Storage and Retrieval]},
  pdfcreator={LaTeX with hyperref package},
  pdfproducer={pdflatex}
}

\usepackage{algpseudocode}
\usepackage{algorithm}
\usepackage{microtype}
\usepackage{booktabs}
\usepackage[table]{xcolor}
\usepackage{graphicx}
\usepackage{subcaption}
 
\usepackage{tikz-qtree}
\usepackage{multirow}
\usepackage{multicol}
\usepackage{amsmath}
\usepackage{mathabx}
\usepackage{pgfplots}
\usetikzlibrary{patterns}
\usepackage{csquotes}
\usepackage{pgfplotstable}
\usepackage{enumitem}
\usepackage{mathrsfs}  
\usepackage{nicefrac}
\usepackage{mathtools}

\newtheorem{mydef}{\textbf{Definition}}
\usepackage[font=small,labelfont=bf]{caption}

\usepackage{xspace} 
\newcommand{\tdsm}{hierarchical significant words language model\xspace}

\newcommand{\TDSM}{Hierarchical Significant Words Language Model\xspace}
\newcommand{\SWLM}{Significant Words Language Model\xspace}
\newcommand{\tdsms}{{\tdsm}s\xspace}

\newcommand{\TDSMs}{{\TDSM}s\xspace}
\newcommand{\actdsm}{HSWLM\xspace}
\newcommand{\ssp}{Strong Separation Principle\xspace}
\newcommand{\acssp}{SSP\xspace}

\newcommand{\shrink}{\vspace{-1.5ex}}
\newcommand{\sshrink}{\vspace{-.75ex}}
\renewcommand{\shrink}{}
\renewcommand{\sshrink}{}

\makeatletter
\DeclareCaptionLabelFormat{numberless}{\ALG@name#1}
\makeatother

\usepackage{times}
\def\:{\hskip0pt} 
\newcommand{\mypar}[1]{\medskip\noindent\textbf{#1}~}
\newcounter{todocnt}

\newcounter{latercnt}

\usepackage[square,comma,numbers,sort&compress,sectionbib]{natbib}
\usepackage{tabularx} 

\def\sharedaffiliation{
\end{tabular}
\begin{tabular}{c}}

\begin{document}
%
\CopyrightYear{2016} 
\setcopyright{acmlicensed}
\conferenceinfo{ICTIR '16,}{September 12 - 16, 2016, Newark, DE, USA}
\isbn{978-1-4503-4497-5/16/09}\acmPrice{\$15.00}
\doi{http://dx.doi.org/10.1145/2970398.2970408}

\title{On Horizontal and Vertical Separation\\ in Hierarchical Text Classification}

%
%
%
%
%

\numberofauthors{1} 
%
\author{
\alignauthor
Mostafa Dehghani\(^1\) 
$\qquad$
Hosein Azarbonyad\(^2\)
$\qquad$
Jaap Kamps\(^1\)
$\qquad$
Maarten Marx\(^2\)\\
\vspace{-10pt}
\sharedaffiliation
\affaddr{\mbox{}\(^1\)Institute for Logic, Language and Computation, University of Amsterdam, The Netherlands}\\
\affaddr{\mbox{}\(^2\)Informatics Institute, University of Amsterdam, The Netherlands}\\
\affaddr{\{dehghani,h.azarbonyad,kamps,maartenmarx\}@uva.nl}
}


\newcommand{\maingoal}{to understand and validate the effect of the separation property on hierarchical classification and discuss how to provide horizontally and vertically separable language models for text\:-\:based hierarchical entities}

\newcommand{\qone}{What makes separability a desirable property for classifiers?}
\newcommand{\findingone}{We demonstrate that based on the ranking and classification principles, \emph{separation property} in the data representation theoretically follows separation in the scores and consequently improves the accuracy of classifiers' decisions. We state this as the ``\ssp'' for optimizing expected effectiveness of classifiers. Furthermore, we define two\:-\:dimensional separation in the hierarchical data and discuss its necessity for hierarchical classification}

\newcommand{\qtwo}{How can we estimate horizontally and vertically separable language models for the hierarchical entities?}
\newcommand{\findingtwo}{ We show that to estimate horizontally and vertically separable language models, they should capture \emph{all}, and \emph{only}, the essential terms of the entities taking their positions in the hierarchy into consideration. Based on this, extending~\citep{Dehghani:2016:CIKM1}, we introduce \TDSMs (\actdsm) and evaluate them on the real\:-\:world data to demonstrate that they provide models for hierarchical entities that possess both horizontal and vertical separability.}

\newcommand{\qthree}{How separability improves transferability?}
\newcommand{\findingthree}{We investigate the effectiveness of language models of hierarchical entities possessing two\:-\:dimensional separation across time and show that separability makes the models capture essential characteristics of a class, which consequently improves transferability over time.}

\maketitle

\begin{abstract}
Hierarchy is a common and effective way of organizing data and representing their relationships at different levels of abstraction. However, hierarchical data dependencies cause difficulties in the estimation of ``separable'' models that can distinguish between the entities in the hierarchy. 
Extracting separable models of hierarchical entities requires us to take their relative position into account and to consider the different types of dependencies in the hierarchy.
In this paper, we present an investigation of the effect of separability in text-based entity classification and argue that in hierarchical classification, a separation property should be established between entities not only in the same layer, but also in different layers. 

Our main findings are the followings.
First, we analyse the importance of separability on the data representation in the task of classification and based on that, we introduce a ``\ssp'' for optimizing expected effectiveness of classifiers decision based on separation property.
Second, we present \TDSMs (\actdsm) which capture all, and only, the essential features of hierarchical entities according to their relative position in the hierarchy resulting in horizontally and vertically separable models.
Third, we validate our claims on real world data and demonstrate that how \actdsm improves the accuracy of classification and how it provides transferable models over time.
Although discussions in this paper focus on the classification problem, the models are applicable to any information access tasks on data that has, or can be mapped to, a  hierarchical structure.
\vspace{-7pt}
\end{abstract}

%
%
%
\keywords{\vspace{-2pt}Separation, Hierarchical Significant Words Language Models, Hierarchical Text Classification\vspace{-5pt}}  





\shrink
\section{Introduction}
\label{sec:int}
\sshrink
Hierarchy is an effective and common way of representing information and many real\:-\:world textual data can be organized in this way.
Organizing data in a hierarchical structure is valuable since it determines relationships in the data at different levels of resolution and picks out different categories relevant to each of the different layers of memberships. 
In a hierarchical structure, a node at any layer could be an indicator of a document, a person, an organization, a category, an ideology, and so on, which we refer to them as ``hierarchical entities''. 
Taking advantage of the structure in the hierarchy requires a proper way for modeling and representing entities, taking their relation in the hierarchy into consideration.

There are two types of dependencies in the hierarchies: i) \emph{Horizontal dependency}, which refers to the relations of entities in the same layer.  A simple example would be the dependency between siblings which have some commonalities in terms of being descendants of the same entity. ii) \emph{Vertical dependency}, which addresses the relations between ancestors and descendants in the hierarchy. For example the relation between root and other entities. 
Due to the existence of two\:-\:dimensional dependencies between entities in the hierarchy, modeling them regardless of their relationships might result in overlapping models that are not capable of making different entities distinguishable. 
Overlap in the models is harmful because when the data representations are not well\:-\:separated, classification and retrieval systems are less likely to work well~\citep{Lewis:1992}. Thus, \emph{two\:-\:dimensional separability}, i.e.\ \emph{horizontal and vertical separability}, is one of the key requirements of hierarchical classification.
\begin{figure*}[t]
\makebox[\linewidth][c]{
\centering
\begin{minipage}[b]{0.30\textwidth} 
\centering
\tikzset{
every tree node/.style={align=center,anchor=north}
edge from parent/.style={very thick},
edge from parent/.style={draw, edge from parent path={(\tikzparentnode.south) -- +(0,-8pt) -| (\tikzchildnode)}},
blank/.style={draw=none}}
\begin{tikzpicture}[level distance=1.5cm, thick,scale=0.76, every node/.style={scale=0.76}]
\matrix
{
\node{\Tree
    [.All  \edge[blank]; 
    [.Status  \edge[blank];
    [.Party \edge[blank]; 
    [.Member ]]]]};
&
\node{\Tree 
    [.Parliament
        [.Government
            [.$P_1$  $M_1$ [.{\dots} ] [.{\dots} ] ]
            [.{\dots} ]
            [.{\dots} ]
        ]
        [.Opposition 
            [.{\dots} ]
            [.{\dots} ]
            [.$P_n$  [.{\dots} ] [.{\dots} ] $M_n$ ]
        ]
    ]};\\
};
\end{tikzpicture}

\vspace{15pt}
\caption{\label{fig:ParHierarchy}Hierarchical relations in the parliament}
\end{minipage}
\vspace{-10pt}
\hspace{\stretch{2}}
\begin{minipage}[b]{0.70\textwidth} 
\begin{subfigure}{0.49\textwidth}
\includegraphics[width=\linewidth]{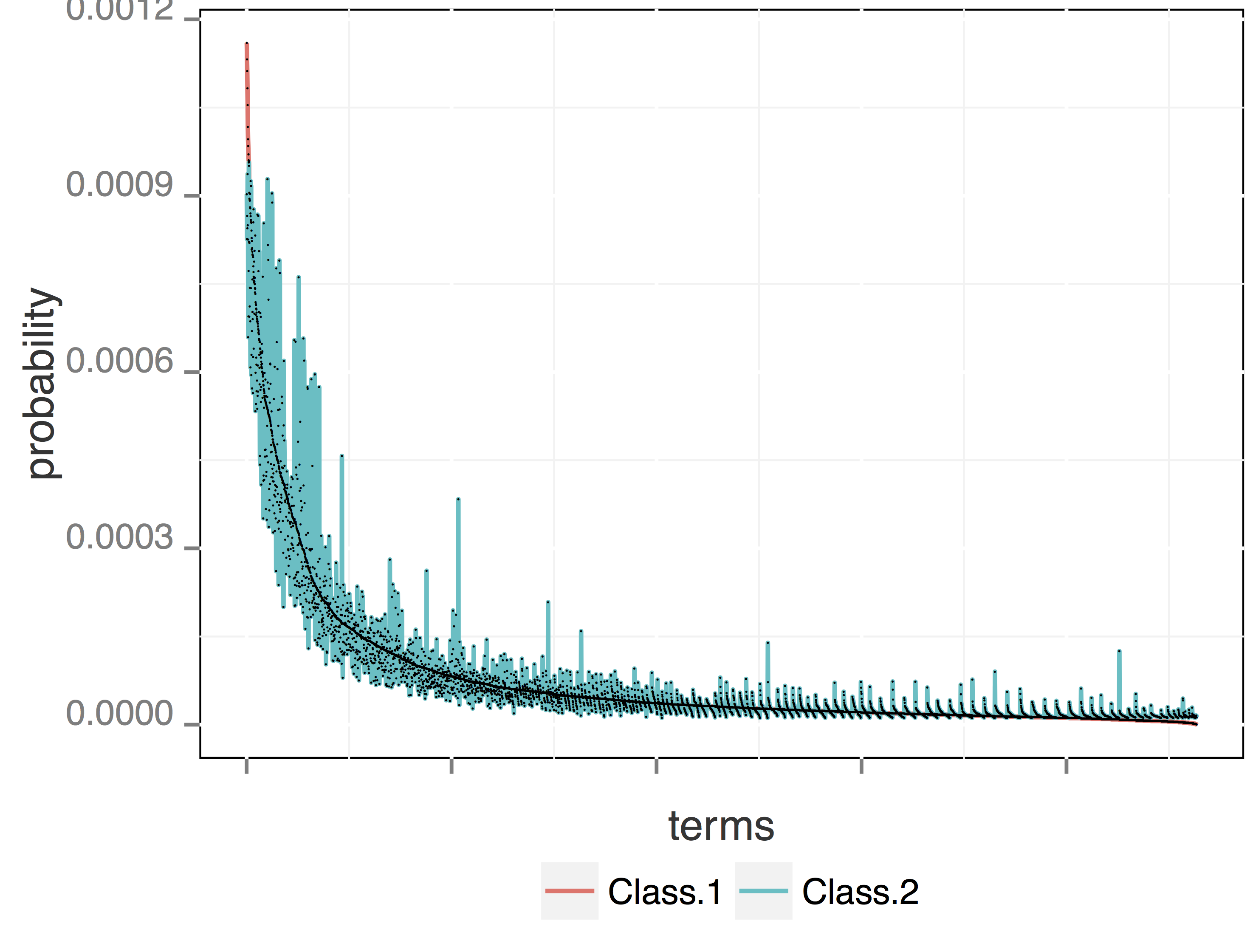}
\caption{\label{fig:sep-slm} \scriptsize{A non\:-\:separable representation of data}}
\vspace{-10pt}
\end{subfigure}
\hfill
\begin{subfigure}{0.49\textwidth}
\includegraphics[width=\linewidth]{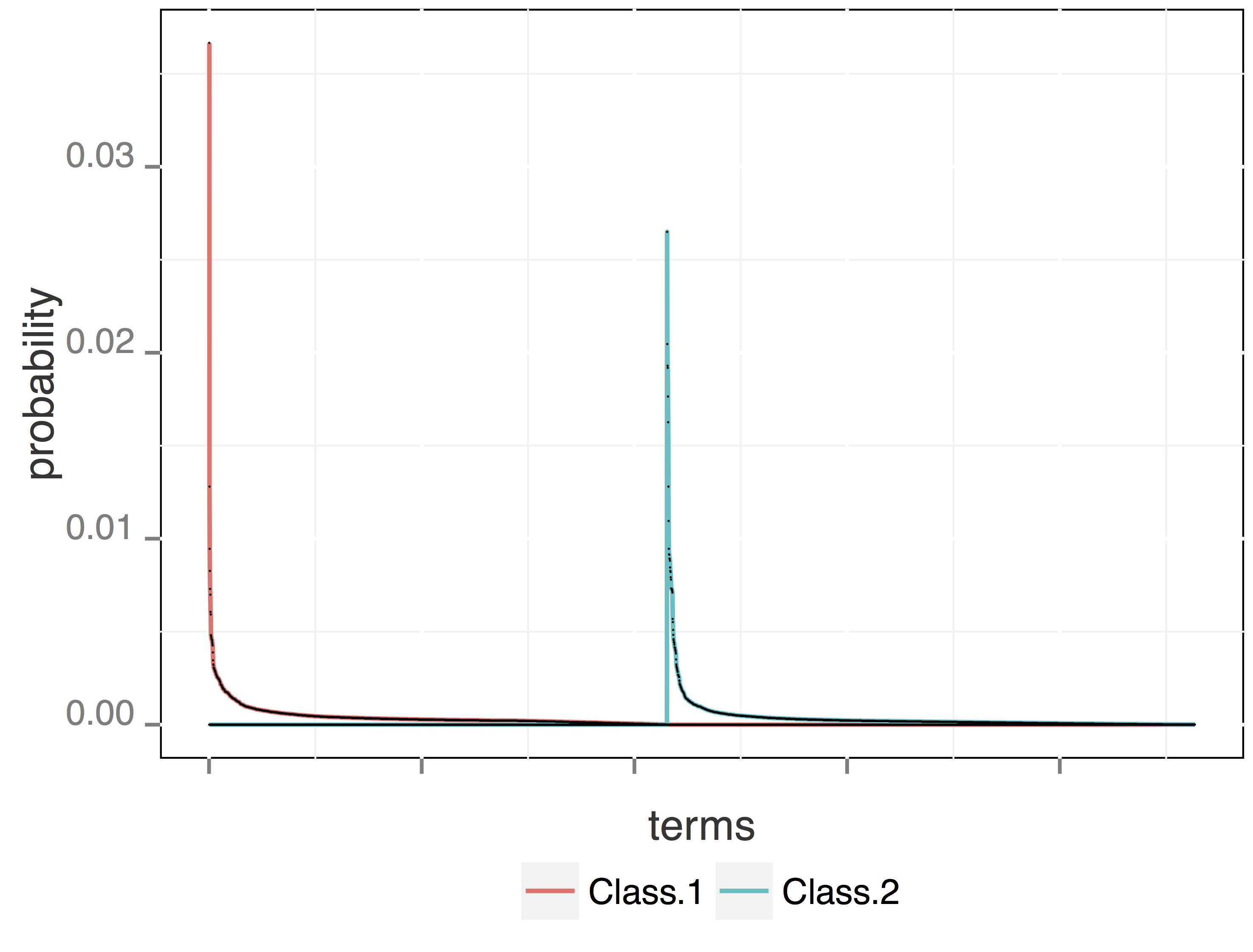}
\caption{\label{fig:sep-glm} \scriptsize{A well\:-\:separable representation of data}}
\vspace{-10pt}
\end{subfigure}
\caption{\label{fig:sep} Probability distribution over terms for data in two different classes, (entities in the statues layer of the parliament), sorted based on the term weights in one of the classes.}
\end{minipage}
\vspace{-10pt}
}
\vspace{-35pt}
\end{figure*}

As a concrete example, consider a simple hierarchy of a multi\:-\:party parliament as shown in Figure~\ref{fig:ParHierarchy}, which determines different categories relevant to the different layers of membership in the parliament.
We can classify these entities based on text, in particular the transcripts of all speeches in parliament as recorded in the parliamentary proceedings.  That is, we can characterize an individual member of parliament by her speeches, a political party by their member's speeches, the opposition by the speeches of members of opposition parties, etc.  
However, in this way, all classifiers are based on speeches of (set of) individual members, making it important to take relations between different layers of the hierarchy explicitly taken into account. 
That is, in order to represent a party in this hierarchy, a proper model would show common characteristics of its members\:---\:not members of other parties (\emph{horizontal} separation), and capture the party's generic characteristics\:---\:not unique aspects of the current members captured in the individual member's layer or aspects of whether the party is in government or opposition captured in the status layer (\emph{vertical} separation).

The concept of separability is of crucial importance in information retrieval, especially when the task is not just ranking of items based on their probability of being relevant, but also making a boolean decision on whether or not an item is relevant, like in information filtering. 
Regarding this concern, \citet{Lewis:1995} has presented the Probability Threshold Principle (PTP), as a stronger version of the Probability Ranking Principle~\citep{Robertson:1977}, for binary classification, which discusses optimizing a threshold for separating items regarding their probability of class membership. 
PTP is a principle based on the separability in the score space. In this paper, we discuss separability in the data representation and define a \emph{\ssp} as the counterpart of PTP in the feature space.


Separation in the data or feature space is a favorable property that not only helps to improve for ranking or classification algorithms, but also brings out characteristic features for human inspection.
Figures~\ref{fig:sep-slm} and ~\ref{fig:sep-glm} illustrate two different ways of modeling two entities in the status layer of the parliamentary hierarchy, i.e., government and opposition. 
Each model is a probability distribution over terms (language model) based on the speeches given by all the members in the corresponding status. In each figure, we sort the terms based on their weights in one of the models, and plot the other in the same order. 
As can be seen, although distributions over terms in Figure~\ref{fig:sep-slm} for two classes are different, they do not suggest highly separable representations for classes. However, estimated language models in Figure~\ref{fig:sep-glm} provide highly separable distributions over terms for two classes, identifying the characteristic terms that uniquely represent each class, and can be directly interpreted.  
Moreover, the language models in Figure~\ref{fig:sep-glm} select a small set of characteristic features, making it easy to learn effective classifiers for classes of interest.
%


The main aim of this paper is \textsl{\maingoal}.
We break this down into three concrete research questions:
\begin{enumerate}
\item[\textbf{RQ1}] \sl\qone
\end{enumerate}
\findingone
\begin{enumerate}
\item[\textbf{RQ2}] \sl\qtwo
\end{enumerate}
\findingtwo
\begin{enumerate}
\item[\textbf{RQ3}] \sl\qthree
\end{enumerate}
\findingthree

The rest of the paper is structured as follows.
Next, in Section~\ref{sec:rel}, we discuss some related work. In Section~\ref{sec:sep}, we argue how separability theoretically improves the accuracy of classification results and discuss horizontal and vertical separation in hierarchical structure. 
Then, we discuss how to estimate \actdsm as two-dimensionally separable models for hierarchical entities in Section~\ref{sec:tdsm}. 
In Section~\ref{sec:exp}, we  analyse separability of \actdsm and provide some experiments to assess the transferability of models using \actdsm. Finally, Section~\ref{sec:con} concludes the paper and suggests some extensions to this research as the future work.

\shrink
\section{Related Work}
\label{sec:rel}
\sshrink
This section discusses briefly the separation property in the related domains and review principles in information retrieval and text classification, which are associated with the concept of separability.  
In addition, some research on classification and topic modeling of hierarchical texts are discussed. 

Separability is a property which makes the data representation sufficient to distinguish instances and consequently enables autonomous systems to easily interpreter the data~\citep{Lewis:1992}. For instance in the classification task, classifiers learn more accurate data boundaries when they are provided with separable representations of data from different classes~\citep{Lewis:1995}. The importance of separability in classifiers has led to the fact that making data separable becomes part of classification. As the most familiar instances, SVM by adding extra dimensions implicitly transform the data into a new space where they are linearly separable~\citep{Burges:1998}. 

Separation is also a pivoting concept in the information retrieval. Separating relevant from non-relevant documents is a fundamental issue in this domain~\citep{Robertson:1977,saracevic:1975,Lavrenko:2001}. In IR, separation plays more important role when instead of giving a rank list, a decision should be made about relevancy of documents, for example in the information filtering task~\citep{Lewis:1992}. As another instance, in the task of relevance feedback, there are some efforts on estimating a distinctive model for relevant documents so that it reflects not only  their similarity, but also their difference from whole collection, i.e., what makes them stand out or separated~\citep{Sparck:2003,Hiemstra:2004,Zhai:SMM:2001}. 

In this paper, we address the separation property in the textual data that are organizable in a hierarchical structure. In a hierarchy, due to the existence of dependencies between entities, estimating separated models is a complex task. There is a range of work on the problem of hierarchical text classification~\citep{Sebastiani:2002, Sun:2001}, which tried to model hierarchical text-based entities. \citet{McCallum:1998} proposed a method for modeling an entity in the hierarchy which tackles the problem of data sparseness in lower layer entities. They used a shrinkage estimator to smooth the model of each leaf entity with the model of its ancestors to make the models more reliable. 
There is also similar research on XML data processing, as hierarchically structured data, which try to incorporate evidence from other layers as the context through mixing each element language models by its parent's models~\citep{sigurbjornsson:2004,ogilvie:2004}.
%

Recently, \citet{Song:2014} tackled the problem of representing hierarchical entities with a lack of training data for the task of hierarchical classification.  In their work, given a collection of instances and a set of hierarchical labels, they tried to embed all entities in a semantic space, then they construct a semantic representation for them to be able to compute meaningful semantic similarity between them.
%

\citet{Zhou:2011} proposed a method that directly tackles the difficulty of modeling similar entities at lower levels of the hierarchy. They used regularization so that the model of lower level entities have the same general properties as their ancestors, in addition to some more specific properties. 
Although these methods tried to model hierarchical texts, but their concerns were not making the models separable. Instead, they mostly addressed the problem of \emph{training data sparseness} \cite{Ha-Thuc:2011,Song:2014,McCallum:1998} or presenting techniques for \emph{handling large scale data} \cite{Gopal:2013,Oh:2011,Xue:2008,Ha-Thuc:2011}.

In terms of modeling hierarchical entities, \citet{Kim:2013} used Hierarchical Dirichlet Process \citep[HDP,][]{Teh:2006} to construct models for entities in the hierarchies using their own models as well as the models of their ancestors.  Also, \citet{Zavitsanos:2011} used HDP to construct the model of entities in a hierarchy employing the models of its descendants. This research tries to bring out precise topic models using structure of the hierarchy, but they do not aim to estimate separable models.  

As we will discuss in Section~\ref{sec:exp}, our proposed approach can be employed as a feature selection method for text classification. Prior research on feature selection for textual information~\citep{SIGIR-Workshop-2010,Forman:2003} tried to improve classification accuracy or computational efficiency, while our method aims to provide a separable representation of data that helps train a transferable model. 
Apart from considering the hierarchical structure, our goals also differ from prior research on transferability of models. For instance, research on constructing dynamic models for data streams~\citep{Yao:2009,Blei:2006} first discovered the topics from data and then tried to efficiently update the models as data changes over the time, while our method aims to identify tiny precise models that are more robust and remain valid over time.  Research on domain adaptation~\citep{Xue:2008:plsa,Chen:2011} also tried to tackle the problem of missing features when very different vocabulary are used in test and training data.  This differs from our approach considering the hierarchical relations, as we aim to estimate separable models that are robust against changes in the structure of entities relations, rather than changes in the corpus vocabulary.

\shrink
\section{Separability in the Hierarchies}
\label{sec:sep}
\sshrink
In this section, we address our first research question: ``\qone''\ 

In addition to the investigation of the separation property in general foundational property of classification and defining a \emph{\ssp}, we discuss a two\:-\:dimensional separation property of hierarchical classification.

\sshrink
\subsection{Separation Property}
Separability is a highly desirable property for constructing and operating autonomous information systems \citep{Lewis:1995}, and especially classifiers. 
Here, we present a step by step argument which shows that based on the classification principles, having better separability in the feature space leads to better accuracy in the classification results.


Based on the \emph{Probability Ranking Principle (PRP)} presented by \citet{Robertson:1977}, \citet{Lewis:1995} has formulated a variant of PRP for binary classification:

\begin{displayquote}
\sshrink
\emph{For a given set of items presented to a binary classification system, there exists a classification of the items such that the probability of class membership for all items assigned to the class is greater than or equal to the probability of class membership for all items not assigned to the class, and the classification has optimal expected effectiveness.}
\sshrink
\end{displayquote}

Since in many applications, autonomous systems need to decide how to classify an individual item in the absence of entire items set, Lewis has extended the PRP to the \emph{Probability Threshold Principle (PTP)}:

\begin{displayquote}
\sshrink
\emph{For a given effectiveness measure,
there exists a threshold $p$, $0<p<1$, such that for any set of items, if all and only those items with probability of class membership greater than $p$ are assigned to the class, the expected effectiveness of the classification will be the best possible for that set of items.}
\sshrink
\end{displayquote}

PTP in fact discusses optimizing the effectiveness of classifiers by making items separable regarding their probability of class membership, which is a discussion on ``separability in the score space''. Based on PTP, optimizing a threshold for separating items is a theoretically trivial task, however, there are practical difficulties. 
%
%
The main difficulty refers to the fact that retrieval models are not necessarily capable of measuring actual probabilities of relevance for documents \citep{Arampatzis:2001}, so they do not  guarantee to generate a set of scores from which the optimum cutoff can be inferred. 
%
%
In this regard, a great deal of work has been done on analysing the score distribution over relevant and non-relevant documents to utilize this information for finding the appropriate threshold between relevant and non-relevant documents \citep{Kanoulas:2009,Arampatzis:2009,Arampatzis:2001}. 
%
%
It is a clear fact that the more the score distributions of relevant and non-relevant documents are separable, the easier it is to determine the optimum threshold. 
So, obtaining the \emph{separation property} in the scores distributions of relevant and non-relevant documents is one of the key focus areas for retrieval models.
%

%
%
%
%
There are two ways to obtain separability in the scores distributions.  We could address the complex underlying process of score generation and investigate ranking functions that yield a separable score distribution, as in the score distributional approaches \citep[e.g.][]{Arampatzis:2001}.  Alternatively, we can investigate ways to provide existing scoring functions with a highly separable representation of the data. 
That is, the ``term distribution'' directly provides information about the ``probability of relevance'' \citep{Crestani:1998} and if there are separable distributions over $terms$ of relevant and non-relevant documents, a scoring function satisfying PRP will generate scores that separate the classes of relevant and non-relevant documents. 
%
%
Thus, a \emph{separation property} on feature distribution for representing the data is a favorable property, which follows better accuracy of classifiers' decisions.

This paper is a first investigation in the role of separation in the term or feature spaces, in which we introduce a formal definition for separability and formulate a principle on the effectiveness of classification based on separation property and leave a more formal treatment to future work.
As a formal and general definition, we can refer to the model separability as follows:
\begin{mydef}
\sshrink
\label{def:sep}
The model of an entity is epistemologically ``{separable}'' if, and only if, it has 
unique, non\:-\:overlapping features that distinguish it from other models.
\sshrink
\end{mydef}

We argued that how separability in feature space leads to the separability in score space. Based on this and the given definition of the separability, we present \emph{\ssp (\acssp)}, which is a counterpart of the PTP~\citep{Lewis:1995} in the feature space:


\begin{displayquote}
\sshrink
\emph{For a given set of items presented to a classification system, for each class there exists at least one feature $\delta$ in the representation of items, and a threshold $\tau$, such that for any set of items, if all and only those items with $\delta > \tau$ are assigned to the class, the classification will have the optimal possible performance for that set of items in terms of a given effectiveness measure.}
\sshrink
\end{displayquote}

\acssp in general is a stronger version of PTP. In strict binary classification, if you have PTP, which holds on the whole feature space, \acssp will be satisfied, however in the multi-class case, \acssp is stronger and it implies PTP, but not the other way around.
Based on PTP, there is always an underlying probabilistic scoring function on the set of \emph{whole} features, which generates membership probabilities as the scores of items and these scores make items separable with regards to a threshold. So, the scoring function can be deemed as a mapping function which maps items to a new feature space in which the score of each item is a single feature representation of that item (membership probabilities (or scores) in PTP would be equivalent to $\delta$ in \acssp). Thus, when the \acssp holds, the PTP and PRP will also hold.  One could envision a stronger version of the \acssp in which ``all'' the features in the representations need to be non-overlapping, but the \acssp is sufficient for optimizing the effectiveness of the classifier.  The separation principle can be formally extended to hierarchical classification in a straightforward way.  In the rest of this section, we will discuss the separation property in the hierarchical classification and explain how to estimate separable representations with the aim of satisfying \acssp in order to improve the classification effectiveness.

\sshrink
\subsection{Horizontal and Vertical Separability}
\label{subsec:vhs}

In hierarchical text classification, there are two types of boundaries existing in the data, horizontal boundaries, and vertical boundaries. Hence, a separation property should be established in two dimensions. It means, not only separation between entities' representation in one layer is required, but also separation between the distribution of terms in different layers is needed.

Separation between entities in the same layer is a related concept to the fundamental goal of all classifiers on the data with flat structure, which is making the data in different classes distinguishable~\citep{Sebastiani:2002}. However, separation between entities in different layers is a concept related to difference of abstraction level and modeling data in different layers in a separable way can help the scoring procedures to figure out the meaning behind the layers and make their decisions less affected by the concepts of other unrelated layers, thus leading to conceptually cleaner and theoretically more accurate models.

Based on Definition~\ref{def:sep}, we formally define horizontal and vertical separability in the hierarchy as follows:
\begin{mydef}
The model of an entity in the hierarchy is ``horizontally separable'' if, and only if, it is \emph{separable} compared to other entities in the same layer, with the same abstraction level.
\end{mydef}

\begin{mydef}
The model of an entity in the hierarchy is ``vertically separable'' if, and only if, it is \emph{separable} compared to other entities in the other layers, with different abstraction levels.
\end{mydef}

To formalize these concepts, consider we have a simple three layers hierarchy of text documents with ``IsA'' relations, where the individual documents take place in the lowest layer, and each node in the middle layer determines a category, representing a group of documents, i.e. its children, and the super node on the top of the hierarchy deemed to represent all the documents in all the groups in the hierarchy. 
There is a key point in this hierarchy to which we will refer for estimating models for the hierarchical entities: ``each node in the hierarchy is a general representation of its descendants''.

First assume that the goal is to estimate a language model representing category $c$, as one of the entities in the middle layer of the hierarchy, and we need the estimated model possessing \emph{horizontal separability}. 
To estimate a horizontally separable model of a category, which represents the category in a way that it is distinguishable from other categories in the middle layer, the key strategy is to eliminate terms that are common across all the categories (overlapping features) and preserve only the discriminating ones.

To do so, we consider there is a general model that captures all the \emph{common} terms of all the categories in the middle layer, $\theta_c^{g}$.  Also we assume that the standard language model of $c$, i.e the model estimated  from concatenation of all the documents in $c$ using MLE, $\theta_{c}$, is drawn from the mixture of the \emph{latent horizontally separable model}, $\theta_{c}^{hs}$, and general model that represents shared terms of all categories, i.e. $\theta_c^{g}$:
\begin{equation}
p(t|\theta_{c})  =  \lambda p(t|\theta_{c}^{hs}) + (1-\lambda) p(t|\theta_c^g),
\end{equation}
where $\lambda$ is the mixture coefficient.
Regarding the meaning of the relations between nodes in the hierarchy, top node in the hierarchy is supposed to be a general representation of all categories. On the other hand $\theta_c^{g}$ supposed to be a model capturing general features of all the categories in the middle layer. Thus, we can approximate $\theta_c^{g}$ with the estimated model of the top node in the hierarchy, $\theta_{all}$:
\begin{equation}
 p(t|\theta_c) \approx \lambda p(t|\theta_{c}^{hs}) + (1-\lambda) p(t|\theta_{all}).
 \label{mixture}
\end{equation}

We estimate $\theta_{all}$ using MLE as follows:
\begin{equation}
p(t|\theta_{all}) = \frac{tf(t,all)}{\sum_{t'} tf(t',all)} = \frac{\sum_{c \in all}{\sum_{d \in c}{tf(t,d)}}}{{\sum_{c \in all}{\sum_{d \in c}{\sum_{t' \in d}{tf(t',d)}}}}},
\label{eq:cmodel}
\end{equation}
where $tf(t,d)$ indicates the frequency of term $t$ in document $d$ and $\theta_{all}$ is in fact collection language model. 

Now, the goal is to extract $\theta_{c}^{hs}$. With regard to the generative models, when a term $t$ is generated using the mixture model in Equation~\ref{mixture}, first a model is chosen based on $\lambda$ and then the term is sampled using the chosen model.  The log-likelihood function for generating the whole category $c$ is:
\begin{equation}
\log p(t|\theta_{c}^{hs}) = \sum_{t \in c} tf(t,c) \log \big(\lambda p(t|\theta_{c}^{hs}) + (1-\lambda) p(t|\theta_{all})\big),
\end{equation}
where $tf(t,c)$ is the frequency of occurrence of term $t$ in category $c$. 
With the goal of maximizing this likelihood function, the maximum likelihood estimation of $p(c|\theta_{c}^{hs})$ can be computed using the Expectation-Maximization (EM) algorithm by iterating over the following steps:
\\
\textbf{E-step:}
\begin{equation}
e_t = tf(t|c).\frac{\lambda p(t|\theta_{c}^{hs})}{\lambda p(x|\theta_{c}^{hs}) + (1-\lambda) p(x|\theta_{all})}
\label{EM-E},
\end{equation}
\textbf{M-step:}
\begin{equation}
p(x|\theta_{c}^{hs}) = \frac{e_t}{\sum_{t' \in \mathcal{V}} e_t'}, \mathrm{~i.e.~normalizing~the~model},
\label{EM-M}
\end{equation}
where $\mathcal{V}$ is the set of all terms with non-zero probability in $\theta_c$. In Equation~\ref{EM-E}, $\theta_c$ is the maximum likelihood estimation of category $c$: $p(t|\theta_{c}) = \nicefrac{\sum_{d \in c}c(t,d)}{\sum_{d \in c}{\sum_{t' \in d} c(t',d)}}$ and $\theta_{c}^{hs}$ represents the horizontally separable model, which in the first iteration it is initialized by the maximum likelihood estimation, similar to $\theta_c$. 
\begin{figure}[!t]
\centering
\begin{algorithm}[H]
 \begin{algorithmic}[1]
 \Procedure{Parsimonize}{$e$,$B$}
 \ForAll{term $t$ in the vocabulary}
 \State \begin{tiny}$P(t|\theta_B) \xleftarrow{normalized} \sum_{b_i\in B} \bigg(P(t|\theta_{b_i}) \prod_{\substack{b_j\in B \\ j \neq i}} (1-P(t|\theta_{b_j}))\bigg)$ \end{tiny}
 \Repeat
 \State \begin{tiny}E-Step: $P[t\in \mathcal{V}] \gets P(t|\theta_e).\frac{\alpha P(t|\tilde{\theta}_e)}{\alpha P(t|\tilde{\theta}_e) + (1-\alpha) P(t|\theta_B)}$ \end{tiny}
  \State \begin{tiny}M-Step: $P(t|\tilde{\theta}_e) \gets \frac{ P[t \in \mathcal{V}]}{\sum_{t' \in \mathcal{V}} P[t' \in \mathcal{V}]}$ \end{tiny}
 \Until{$\tilde{\theta_t}$ becomes stable}
 \EndFor
 \EndProcedure
 \end{algorithmic}
 \caption{Modified Model Parsimonization}
\end{algorithm}
\vspace{-10pt}
\caption{\label{Fig:Alg:PARS}Pseudo-code for procedure of modified model parsimonization.}
\vspace{-15pt}
\end{figure}

Considering the above process, a horizontally separable model is a model which is \textbf{specified} by taking out general features that have high probability in ``all'' categories, or lets say collection language model, which is similar to the concept of the parsimonious language model, introduced by~\citet{Hiemstra:2004}.

Now assume that we want to extract a language model possessing \emph{vertical separability} for the category $c$, i.e. a model that makes this category distinguishable from entities both in the lower layer (each individual document) and the top layer (collection of all documents).
In the procedure of making the model horizontally separable, we argued that we can reduce the problem to removing terms representing the top node, which results in a model that is separable from the top node in the upper layer. This means that we are already half-way towards making a model vertically separable, thus, the model only requires to be separable from its descendant entities in the lower layer.  
Back to the meaning of the ``IsA'' relations in the hierarchy, the model of each node is a general representation of all its descendants. So making the model of a category separable from its descendant documents means to remove terms that describe each individual documents, but not all of them. We call these terms, document specific terms.
\begin{figure}[!t]
\centering
\begin{algorithm}[H]
 \begin{algorithmic}[1]
 \Procedure{Estimate{\actdsm}s}{}
 \Statex ~~~~Initialization: \vspace{2pt}
 \ForAll{entity $e$ in the hierarchy}
 \State $\theta_e \gets$ standard estimation for $e$ using MLE
 \EndFor
 \Repeat
 \State \Call{Specification}{}
 \State \Call{Generalization}{} 
 \Until{models do not change significantly anymore}
 \EndProcedure
 \end{algorithmic}
 \caption{Estimating \TDSMs}
\end{algorithm}
\vspace{-10pt}
\caption{\label{Fig:Alg:IDSP}Pseudo-code for the overall procedure of estimating \actdsm.}
\vspace{-15pt}
\end{figure}
For each category $c$, we assume there is a  model, $\theta_d^{s}$, that captures document specific terms, i.e. terms from documents in that category that are good indicators for individual documents but not supported by all of them. Also we assume that the standard language model of $c$, $\theta_{c}$, is drawn from the mixture of the \emph{latent vertically separable model}, $\theta_{c}^{vs}$, and $\theta_d^{s}$: 
\begin{eqnarray}
p(t|\theta_{c}) & = & \lambda p(t|\theta_{c}^{vs}) + (1-\lambda) p(t|\theta_d^s)
\end{eqnarray}
where $\lambda$ is the mixing coefficient. 
We estimate $\theta_d^s$ using the following equation:
\begin{equation}
p(t|\theta_d^s)  \xleftarrow{normalized} 
\sum_{d_i\in c} 
\bigg(
p(t|\theta_{d_i}) \prod_{\substack{d_j\in c \\ j \neq i}} (1-p(t|\theta_{d_j}))
\bigg),
\label{eq:smodel}
\end{equation}
where $p(t|\theta_{d_i}) = \nicefrac{c(t,d_i)}{\sum_{t' \in d_i} c(t',d_i)}$. This equation assigns a high probability to a term if it has high probability in one of the document models, but not others, marginalizing over all the document models.  This way, the higher the probability is, the more specific the term will be. 
Now, the goal is to extract the $\theta_{c}^{vs}$. An EM algorithm, similar to Equations~\ref{EM-E} and~\ref{EM-M} can be applied for estimating $\theta_{c}^{vs}$ by removing the effect of $\theta_d^s$ from $\theta_{c}$.

Considering the above process, a vertically separable model is a model which is \textbf{generalized} by taking out specific terms that have high probability in one of the descendant documents, but not others.  

\subsection{Two\:-\:Dimensional Separability}
In order to have fully separable models in hierarchical classification, they should own two\:-\:dimensional separation property. We define two\:-\:dimensional separability as follows:
\begin{mydef}
\sshrink
The model of an entity in the hierarchy is ``two\:-\:dimensionally separable'' if, and only if, it is both horizontally and vertically separable at the same time.
\sshrink
\end{mydef}

Intuitively, if a model of an entity is two\:-\:dimensionally separable, it should capture \emph{all}, and \emph{only}, the essential features of the entity taking its relative position in the hierarchy in to consideration.  In the next section, we will discuss how to estimate two\:-\:dimensional separable models for entities in the hierarchies with more than three layers.

\smallskip
In summary, 
based on the discussions on this section, we can say that the separation property is a desirable foundational property for classifiers. We see that based on PRP, separation in the feature space follows by separation in the score space, which leads to improvement in classification accuracy.  We also notice that separation property in hierarchies is defined in two dimensions, thus, fully separable hierarchical models would possess both horizontal and vertical separation.  

\shrink
\section{\TDSMs}
\label{sec:tdsm}
\sshrink
\begin{figure*}[t]
\renewcommand{\thesubfigure}{a}
\begin{subfigure}[t]{0.45\textwidth}
\centering
\begin{algorithm}[H]
\captionsetup{labelformat=empty}
\begin{algorithmic}[1]
\Procedure{Specification}{}
\State Queue $\gets$ all entities in breadth first order
\While{Queue is not empty}
  \State $e \gets$ Queue.pop()
  \State $l \gets e$.Depth()
  \While{$l > 0$}
    \State $A \gets e$.\Call{getAncestor}{$l$}
    \State \Call{Parsimonize}{$e$,$A$}
    \State $l \gets l-1$  
  \EndWhile
\EndWhile
\EndProcedure
\end{algorithmic}
\captionof{algorithm}{Specification Stage}
\end{algorithm}
\vspace{-10pt}
\captionof{figure}{\label{Fig:Alg:SpecStage}Procedure of Specification. $e$.\Call{getAncestor}{$l$} gives the ancestor of entity $e$ with $l$ edges distance from it.}
\end{subfigure}
~~~~~~~~~~~~~~~~
\renewcommand{\thesubfigure}{b}
\begin{subfigure}[t]{0.45\textwidth}
\centering
\begin{algorithm}[H]
\captionsetup{labelformat=empty}
\begin{algorithmic}[1]
\Procedure{Generalization}{}
\State Stack $\gets$ all entities in breadth first order
\While{Stack is not empty}
  \State $e \gets$ Stack.pop()
  \State $l \gets e$.Height()
  \While{$l > 0$}
    \State $D \gets e$.\Call{getDecedents}{$l$}
    \State \Call{Parsimonize}{$e$,$D$}
    \State $l \gets l-1$  
  \EndWhile
\EndWhile
\EndProcedure
\end{algorithmic}
\captionof{algorithm}{Generalization Stage}
\end{algorithm}
\vspace{-10pt}
\captionof{figure}{\label{Fig:Alg:GeneStage}Procedure of Generalization. $e$.\Call{getDecedents}{$l$} gives all the decedents of entity $e$ with $l$ edges distance from it.}
\end{subfigure}
\vspace{-10pt}
\caption{\label{IDSP}Pseudo-code for stages of estimating \actdsm. Function \textsc{Parsimonize}($e$,$B$)  parsimonizes $\theta_e$ toward background models in $B$}
\vspace{-15pt}
\end{figure*}

In this section, we address our second research question: ``\qtwo''\
We introduce \TDSMs (\actdsm) which is an extension of \SWLM proposed by~\citet{Dehghani:2016:CIKM1} to be applicable for hierarchical data. \actdsm is in fact a particular arrangement of multiple passes of the procedures of making hierarchical entities' models vertically and horizontally separable, as they are explained in Section~\ref{subsec:vhs}.
Generally speaking, \tdsms are an extension of parsimonious language models \cite{Hiemstra:2004} tailored to text\:-\:based hierarchical entities.  
In parsimonious language model, given a raw probabilistic estimation, the goal is to re-estimate the model so that non-essential parameters of the raw estimation are eliminated with regard to the background estimation. 
The proposed approach for estimating \tdsm iteratively reestimates the standard language models of entities to minimize their overlap by discarding non\:-\:essential terms from them. 

In the original parsimonious language model~\citep{Hiemstra:2004}, background model is explained by the estimation of the \emph{collection model}, i.e. the model representing all the entities, similar to Equation~\ref{eq:cmodel}.  
However, with respect to the hierarchical structure, and our goal in \actdsm for making the entities' models separable from each other, we need to use parsimonization technique in different situations: 1) toward ancestors of an entity, and 2) toward its descendants. Hence, beside parsimonizing toward a single parent entity in the upper layers, as the background model, we need to be  able to do parsimonization toward multiple descendants in the lower layers. 
Figure~\ref{Fig:Alg:PARS} presents pseudo-code of  Expectation-Maximization algorithm which is employed in the modified model parsimonization procedure. 
In the equation in line 3 of the pseudo-code in Figure~\ref{Fig:Alg:PARS}, $B$ is the set of background entities\:---\:either one or multiple, and $\theta_{b_i}$ demonstrates the model of each background entity, $b_i$, which is estimated using MLE. As can be seen, in case of having a single ancestor node as the background model,  this equation will be equal to Equation~\ref{eq:cmodel} and in case of having  multiple descendants as the background models, it results same as Equation~\ref{eq:smodel}. 
In this procedure, in general, in the E-step, the probabilities of terms are adjusted repeatedly and in the M-step, adjusted probability of terms are normalized to form a distribution. 
Another change in the modified version of model parsimonization, which practically makes no difference in the final estimation, is that in the E-step, instead of using $tf(t,e)$, we employ $p(t|\theta_e)$, where $\theta_e$ is the model represents entity $e$ and initially it is estimated using MLE. This is because in the multi-layer hierarchies, there are more than one parsimonization pass for a particular entity and after the first round, we need to use the probability of terms estimated from the previous pass, not the raw information of their frequency.

Model parsimonization is an almost parameter free process. The only parameter is the standard smoothing parameter $\lambda$, which controls the level of parsimonization, so that the lower values of $\lambda$ result in more parsimonious models.
The iteration is repeated a fixed number of times or until the estimates do not change significantly anymore. 

The pseudo-code of overall procedure of estimating \actdsm is presented in Figure~\ref{Fig:Alg:IDSP}. 
Before the first round of the procedure, a standard estimation like maximum likelihood estimation is used to construct the initial model for each entity in the hierarchy. 
Then, models will be updated in an iterative process until all the estimated models of entities become stable. 
In each iteration, there are two main stages: a \emph{Specification stage} and a \emph{Generalization stage}. 
In these stages, language models of entities in the hierarchy are iteratively made ``specific,'' by taking out terms explained at higher levels, and ``general,''  by eliminating specific terms of lower layers, which results in models that are both \emph{horizontally} and \emph{vertically} separable as it is described in Section~\ref{subsec:vhs}.

In the specification stage, the goal is to eliminate the general terms of the language model of each entity so that the resulted language model demonstrates entity's specific properties.  
To do so, the parsimonization method is used to parsimonize the language model of an entity towards its ancestors, from the root of the hierarchy to its direct parent, as the background estimations. 
The order in the hierarchy is of crucial importance here. 
When a language  model of an ancestor is considered as the background language model, it should demonstrate the ``specific'' properties of that ancestor. Due to this fact, it is important that before considering the language model of an entity as background estimation, it has already passed the specification stage, and we have to move top-down.
Pseudo\:-\:code of the recursive procedure of specification of entities' models in the hierarchy is depicted in Figure \ref{Fig:Alg:SpecStage}.

In the generalization stage, the goal is to refine language models by removing terms which do not address the concepts in the level of abstraction of the entity's layer.
To do so, again parsimonization is exploited but towards descendants, which leads to elimination of specific terms. 
Here also, before considering the model of an entity as the background estimation, it should be already passed the generalization stage, so generalization moves bottom up.
Figure \ref{Fig:Alg:GeneStage} presents the pseudo\:-\:code for the recursive procedure of generalization of entities' language models in the hierarchy. 
In the generalization step, the background models of descendants are supposed to be specific enough to show their extremely specific properties. Hence, generalization stages must be applied on the output models of specification stages: specification should precede generalization, as shown in Figure \ref{Fig:Alg:IDSP} before.

\smallskip
In this section, we explained the procedure of estimating \tdsm, which terminates in the language models that capture \emph{all}, and \emph{only} the essential terms regarding the hierarchical positions of entities. We will investigate the effectiveness of \actdsm on the real data in the next section.

\shrink
\section{Experiments}
\label{sec:exp}
\sshrink
In order to evaluate the separability of \TDSMs, we use parliamentary data as one of the interesting collections with hierarchically structured data, using the hierarchical entities as shown in Figure~\ref{fig:ParHierarchy}.  First we introduce the collection we have used and then we analyse the quality of \actdsm on providing horizontal and vertical separability over the hierarchy. 

\sshrink
\subsection{Experimental Dataset}
\label{subsec:dataset}

\begin{figure}[!t]
\centering
\makebox[\linewidth]{
\tikzset{
every tree node/.style={align=center,anchor=north}
edge from parent/.style={very thick},
blank/.style={draw=none}}
\begin{tikzpicture}
\begin{small}
\matrix
{
\node{
    \Tree 
    [.Parliament
        [.Government 
            [.$VVD$  $43$ ]
            [.$PvdA$  $42$ ]
        ]
        [.Opposition 
            [.$CDA$  $13$ ]
            [.$PVV$  $12$ ]
            [.$SP$  $18$ ]
            [.$D66$  $12$ ]
            [.$CU$  $5$ ]
            [.$GL$  $4$ ]
        ]
    ]};\\
};
\end{small}
\end{tikzpicture}
}
\vspace{-20pt}
\caption{Composition of house of representatives of Dutch parliament, 2012-2014. \emph{VVD}:People's Party for Freedom and democracy, \emph{PvdA}:Labour Party, \emph{CDA}:Christian Democratic Appeal, \emph{PVV}:Party for Freedom, \emph{SP}:The Socialist Party, \emph{D66}:Democrats 66, \emph{GL}:Green-Left, \emph{CU}:Christian-Union}
\label{fig:DutchParl}
\vspace{-25pt}
\end{figure}
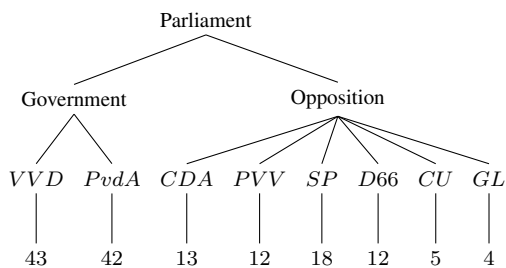

We have made use of the Dutch parliamentary data to do a number of experiments. 
%
As a brief background, the Dutch parliament is a bicameral parliament which consists of the senate and the house of representatives. The house of representatives is the main chamber of parliament, where discussion of proposed legislation and review of the government's actions takes place.  The Dutch parliamentary system is a multi-party system, requiring a coalition of parties to form the government~\citep{deswaan73}.
For the experiments and analysis, we have used the data from the House of Representatives of Netherlands, consisting of literal transcripts of all speeches in parliament with rich annotation of the speaker and debate structure.  We have chosen the last periods of parliament where eight main parties have about 95 percent of the 150 seats in the parliament.  This data collected from March 2012 to April 2014 and consist of 62,206 debates containing 3.7M words. Figure~\ref{fig:DutchParl} shows the hierarchical structure of house of representatives in this period.
For each member, all speeches are collected from the discussions in the parliament and members for which the length of all their given speeches is less than 100 words are removed from the data instances.  No stemming and no lemmatization is done on the data and also stop words and common words are not removed in data preprocessing.   

\sshrink
\subsection{Two\:-\:Dimensional Separability of \actdsm}
\label{subsec:actdsmep}

In this section we investigate the ability of \actdsm on providing language models for hierarchical entities that are two\:-\:dimensionally separable. 
Based on the explained procedure of estimating \actdsm, the language models of entities in the hierarchy is repeatedly updated, so that the resulting models are both \emph{horizontally} and \emph{vertically} separable in the hierarchy. In order to assess this fact, we estimate \actdsm on the parliamentary data and look into the separability between entities in the same layer or in different layers.

\begin{figure*}[!t]

        \centering
        \begin{subfigure}[b]{0.32\textwidth}
\includegraphics[width=\linewidth]{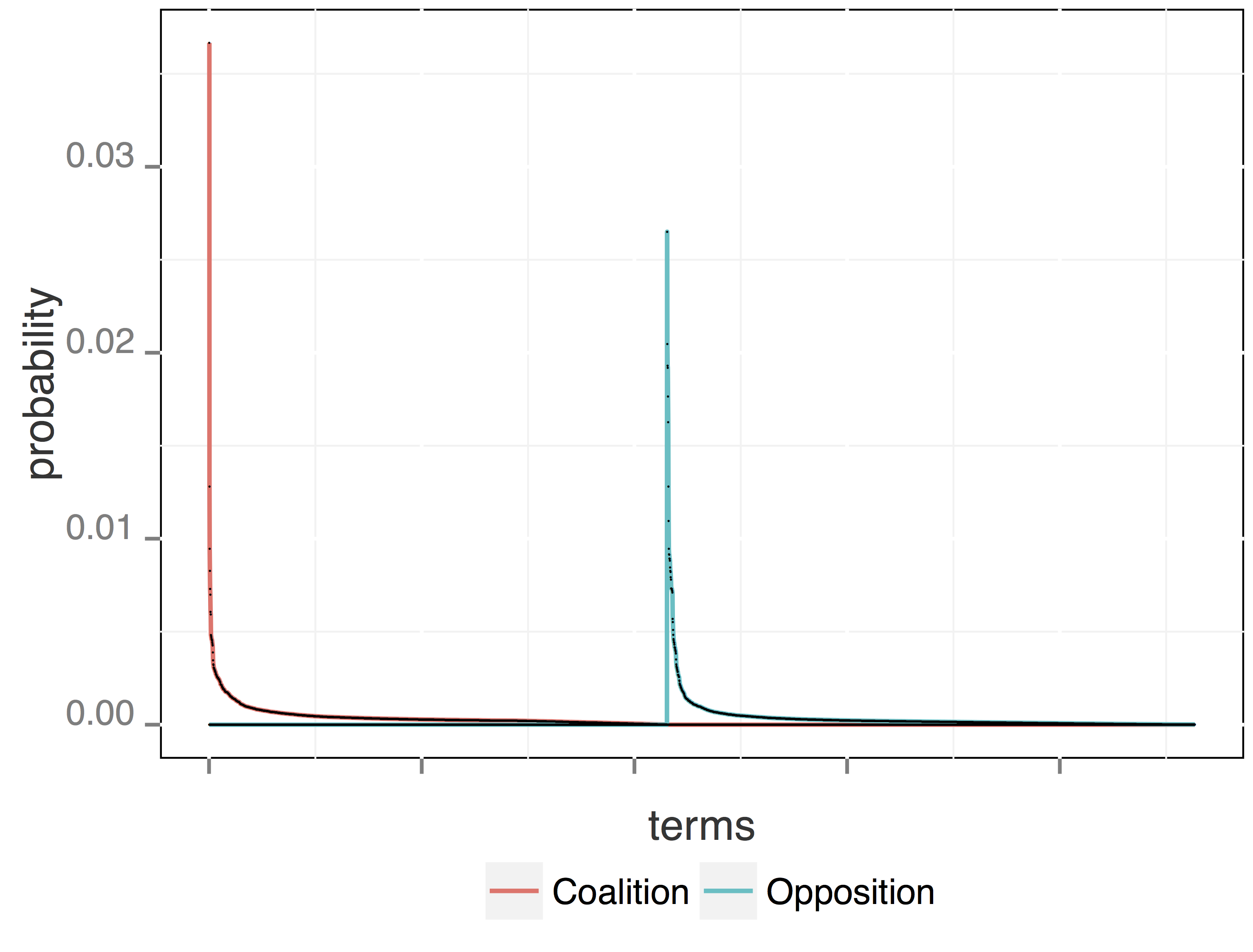}
\caption{\label{fig:HSS}\actdsm in the status layer}
        
        \end{subfigure}
        ~ 
                \begin{subfigure}[b]{0.64\textwidth}
\includegraphics[width=\linewidth]{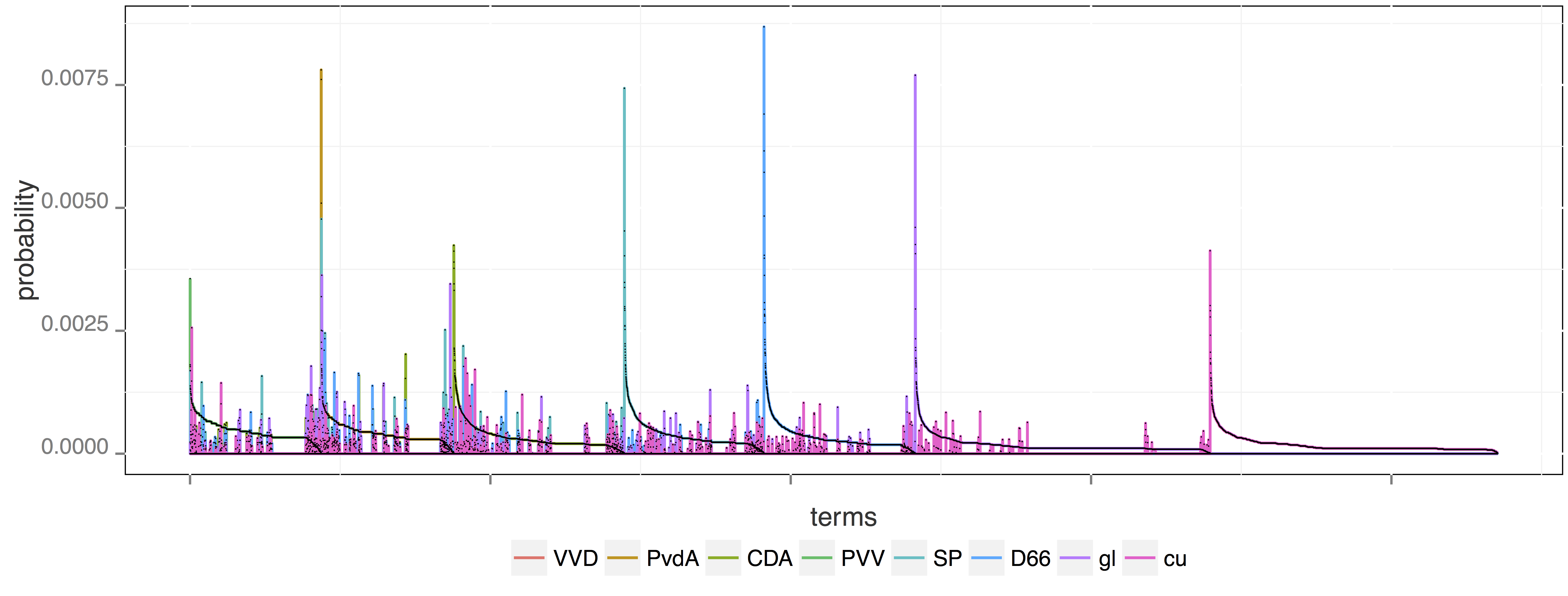}
\caption{\label{fig:HSP}\actdsm in the party layer}
        
        \end{subfigure}
        
        \caption{\label{fig:HS} \emph{Horizontal Separability}: probability distribution over terms based on \tdsms in status layer and party layer.}

\end{figure*}

Figures~\ref{fig:HSS} and~\ref{fig:HSP} illustrate the probability distribution over terms based on the estimated \actdsm in the status and party layer respectively. We sort the probability distribution on the term weight of the first model, and plot the other models in this exact order.
As can be seen in the status layer, Figures~\ref{fig:HSS}, the distributions over terms for government and opposition cover almost separated set of terms. Since in this layer these two entities are supposed to be against each other, a high level of separability can be expected. On the other hand, in the party layer, Figures \ref{fig:HSP}, it is possible that two parties share some ideological issues and consequently share some terms. So, in this layer a complete separability of terms would not be practically possible for all the parties. Nevertheless, \actdsm provides an acceptable horizontal separability in this layer.

\begin{figure*}[!t]
        \centering
        \begin{subfigure}[b]{0.32\textwidth}
        
\includegraphics[width=\linewidth]{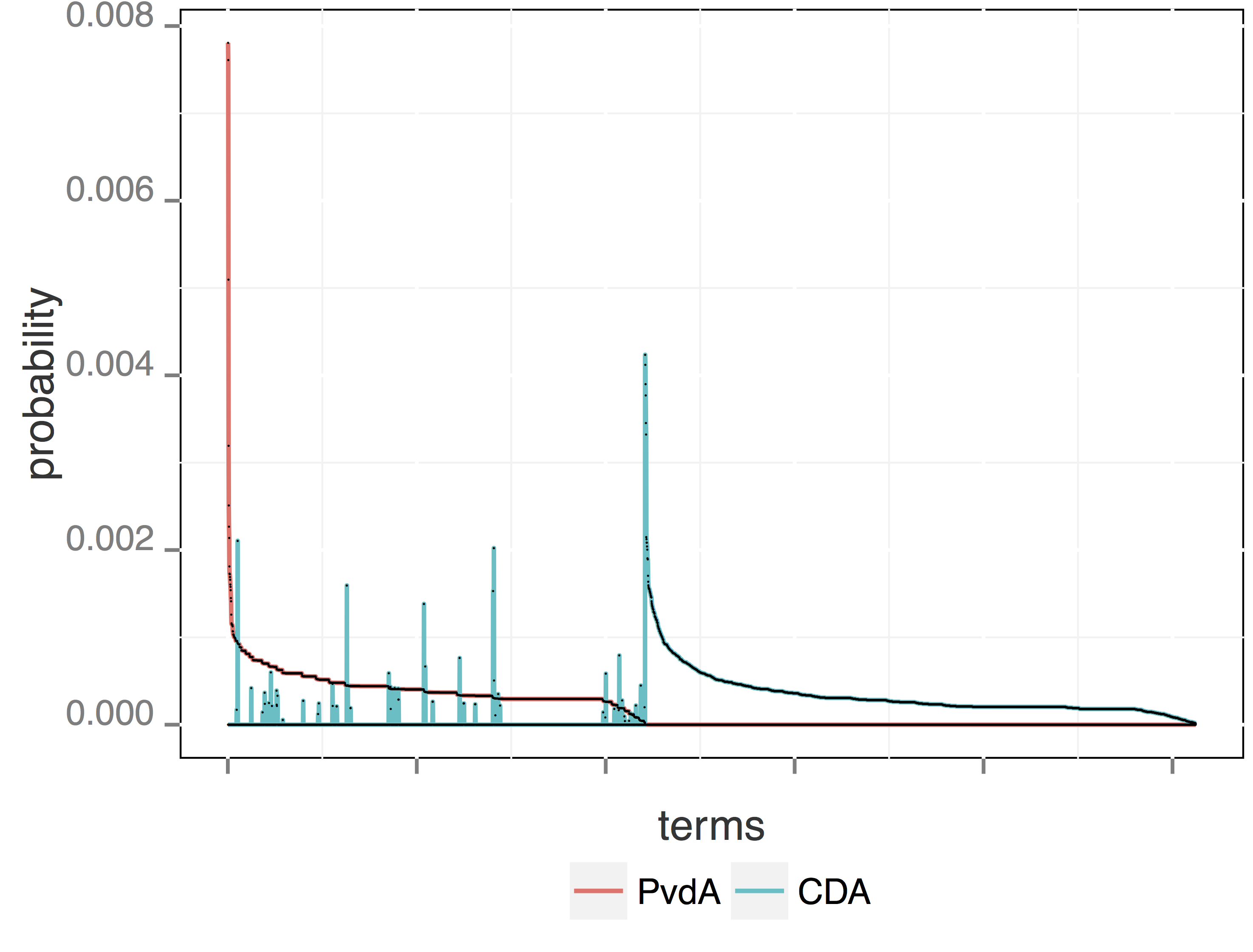}
\caption{\label{fig:HSPCO} \actdsm of two parties in different statuses: Christian Democratic Appeal (CDA) and Labour Party (PvdA)}

        \end{subfigure}
        ~ 
        \begin{subfigure}[b]{0.32\textwidth}
           
\centering
\includegraphics[width=\linewidth]{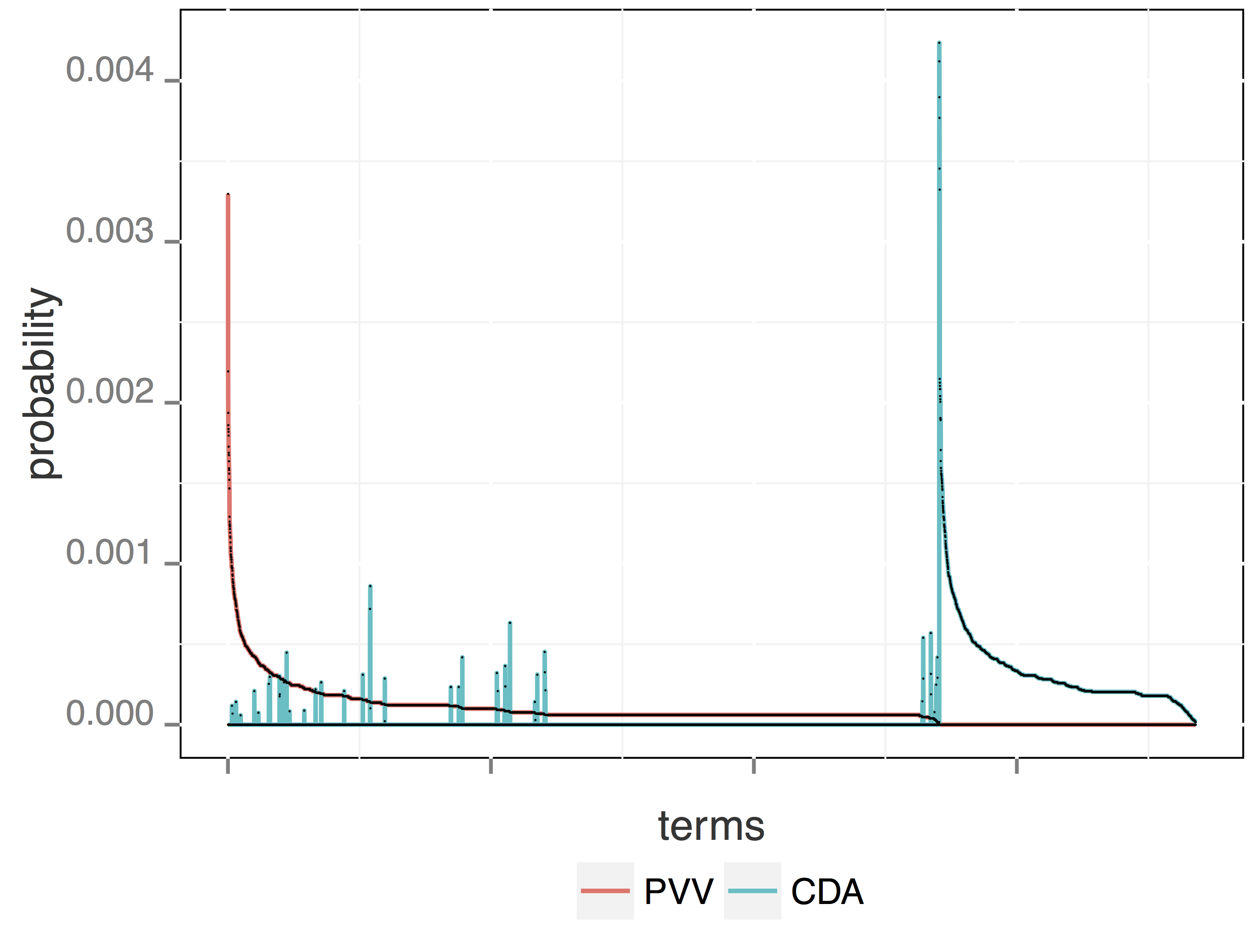}
\caption{\label{fig:HSPOO} \actdsm of two parties in opposition: Party for Freedom (PVV) and Christian Democratic Appeal (CDA)}

        \end{subfigure}
        ~ 
        \begin{subfigure}[b]{0.32\textwidth}
\centering
\includegraphics[width=\linewidth]{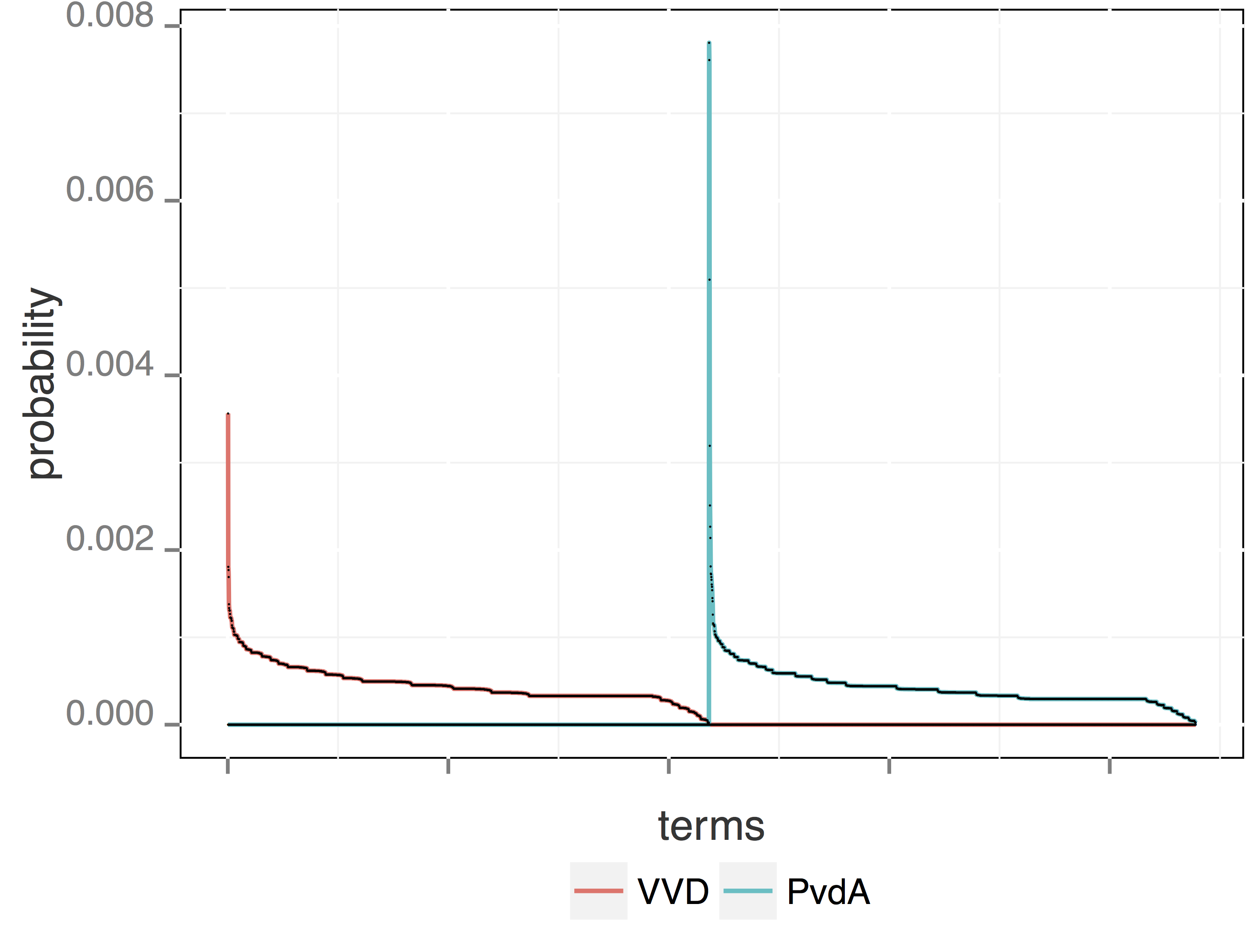}
\caption{\label{fig:HSPCC} \actdsm of two parties in government: People's Party for Freedom (VVD) and Labour Party (PvdA)}
        \end{subfigure}
        
        \caption{\label{fig:HSP-pairs} \emph{Horizontal Separability}: probability distribution over terms based on \tdsms in party layer}
\end{figure*}

In addition, we illustrate the horizontal separability of \actdsm of some pairs of parties. Figures \ref{fig:HSPCO}, \ref{fig:HSPOO}, and \ref{fig:HSPCC} show the separability of models of two parties in three cases, respectively: 1) different statuses, 2) both in the status of opposition, 3) both in the status of government. It can be seen that in all cases of being in the same status or different status, estimated \tdsms are separable. The interesting point is in Figure~\ref{fig:HSPCC} that presents the models of two government parties that are strongly separable. This rooted in the fact that in this period there was an unusual coalition government consisting of a right\:-\:wing and a left\:-\:wing party. So, although they have agreement in the status layer, their model is highly separable in terms of having opposite spectrum in party layer.

\begin{figure*}[!t]
        \centering
        \begin{subfigure}[b]{0.45\textwidth}
\includegraphics[width=\linewidth]{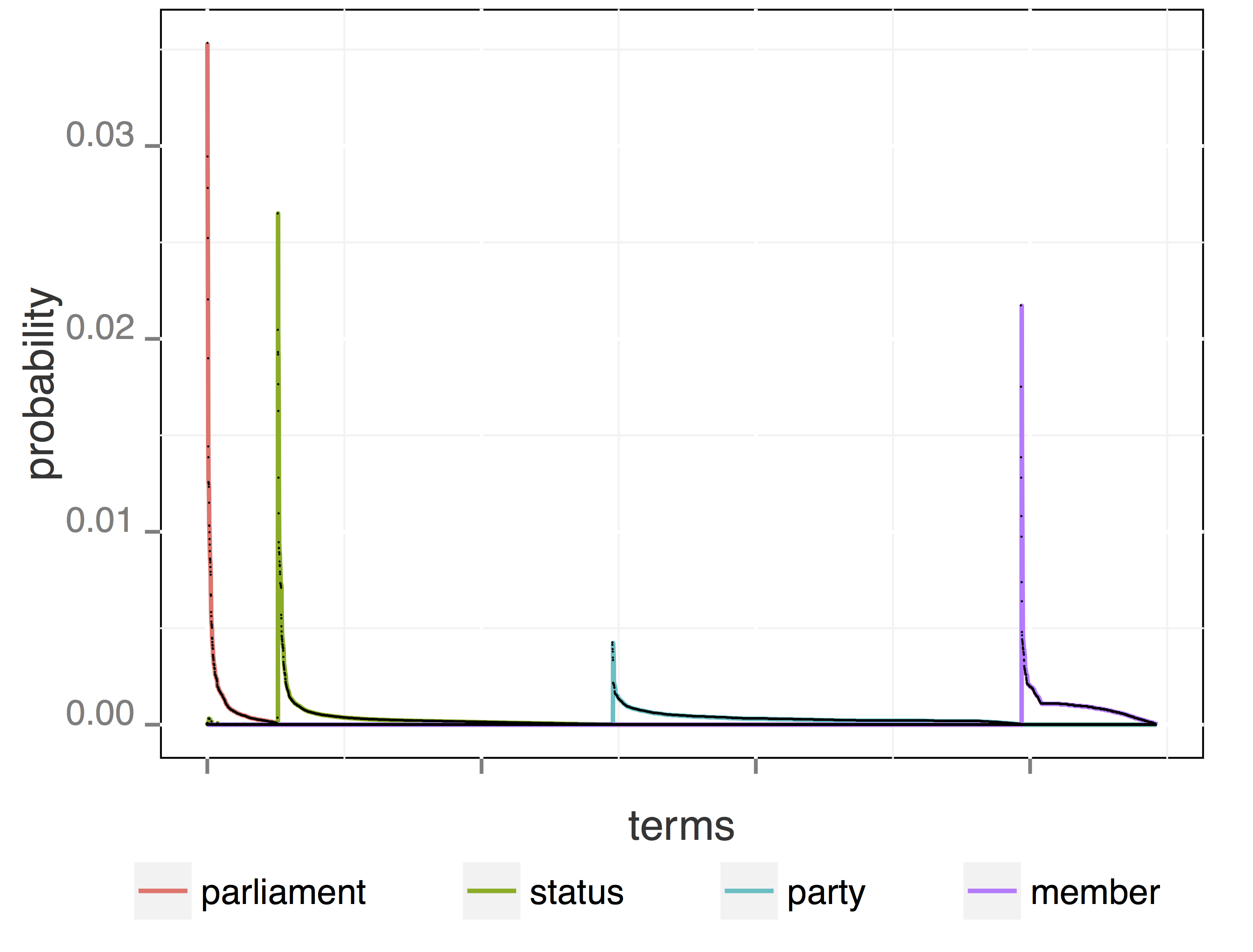}
\caption{\label{fig:VSO} \actdsm of S. van Haersma Buma (as the member of parliament - Leader of CDA), Christian Democratic Appeal (as the party), Opposition (as the status), and the Parliament}
        \end{subfigure}
        ~~~~~~~~
        \begin{subfigure}[b]{0.45\textwidth}
        
\centering
\includegraphics[width=\linewidth]{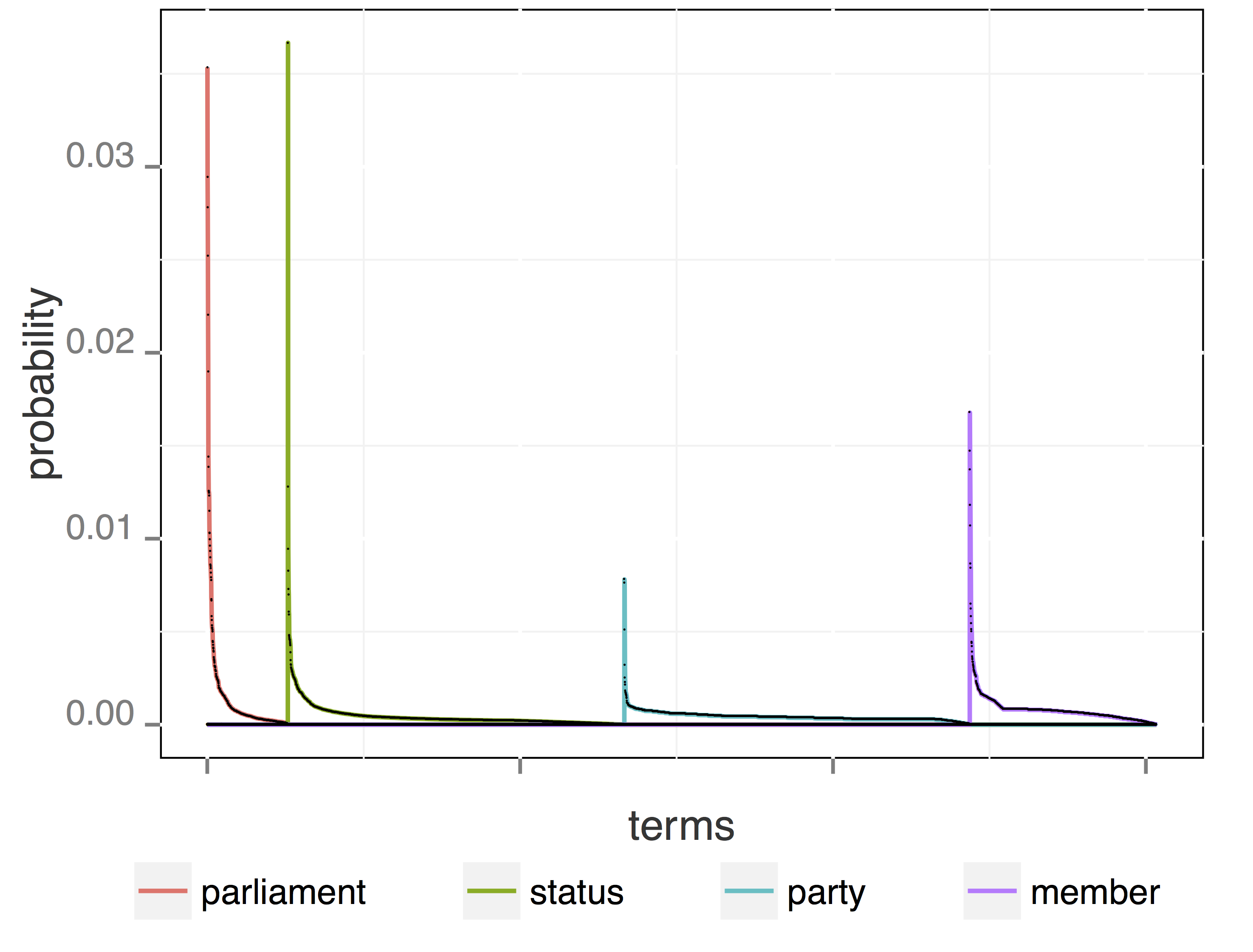}
\caption{\label{fig:VSC} \actdsm of D. Samson (as the member of parliament - Leader of PvdA), Labour Party (as the party), Government (as the status), and the Parliament}

        \end{subfigure}
        
        \caption{\label{fig:VS} \emph{Vertical Separability}: probability distribution over terms in different layers based on \tdsms in complete paths from the root to the terminal entities in the hierarchy}

\end{figure*}

In order to illustrate the vertical separability of \actdsm, we choose two different branches in the hierarchy: one from leader of one of the opposition parties to the root, and the other from leader of one of the government parties to the root. Figures \ref{fig:VSO} and \ref{fig:VSC} show probability distributions over words based on \actdsm of all entities in these two branches. They demonstrate that using \actdsm, we can decompose distribution over all terms to the highly separable distributions, each one representing the language usage related to the meaning behind the layer of the entity in the hierarchy. 

Two\:-\:dimensional separation property of \actdsm in the hierarchy is essentially due to the parsimonization effect in two directions. 
Intuitively, the horizontal separability is mainly the result of specification stage. For example, when an entity is parsimonized toward its direct parent, since the data in its parent is formed by pooling the data from the entity and its siblings, parsimonization makes the model of the entity separable from its siblings, which provide \emph{horizontal separation} in the resulting language models. On the other hand, vertical separability is mainly due to generalization stage (and implicitly specification). For example, when an entity is parsimonized towards its children, since they are specified already, parsimonization gets rid of the specific terms of the lower layer from the entity's model.

\sshrink
\subsection{Separability for Transferability}
\label{subsec:Separability}

Here, we address our third research question: ``\qthree''\ 

To address this question, we investigate the effectiveness of the models in the cross period classification task in the parliamentary dataset, which is to predict the party that a  member of the parliament belongs to, having all the speeches given by that member in a period,  as well as all the speeches given by the members of all parties in a different period of parliament.
In the parliament, the status of the parties may change over different periods. Since the speeches given by the members are considerably affected by the status of their party, a dramatic change may happen in the parties' language usage. Due to this fact, learning a transferable model for party classification over periods is a very challenging task \citep{Hirst:2014,yu:2008}.  

\newcolumntype{Y}{>{\centering\arraybackslash}X}
\begin{table*}[t]
\captionof{table}{\label{tbl:party} Results of party classification task in terms of macro-average accuracy. We have conducted paired t-test to investigate statistical significance of the improvements of the best method over the second best method, in the corresponding experiments. Improvements that are annotated with $^\blacktriangleup$ are statistically significant with p\:-\:value $<$ 0.005.}
\vspace{-5pt}
\begin{tabularx}{\textwidth}{c Y Y Y Y Y Y Y}
\toprule
\multicolumn{2}{c}{} & \multicolumn{6}{c}{\footnotesize Test}
\\ \cmidrule(lr){3-8} 
\multicolumn{2}{c}{} & \multicolumn{2}{c}{\textbf{$SVM$}} & \multicolumn{2}{c}{\textbf{$SVM_{IG}$}} & \multicolumn{2}{c}{\textbf{$SVM_{\actdsm}$}}
\\  \cmidrule(lr){3-4} \cmidrule(lr){5-6} \cmidrule(lr){7-8}
\multicolumn{1}{c}{} & \multicolumn{1}{c}{\multirow{1}{*}{\textbf{Period}}} &2010-2012&2012-2014 & 2010-2012&2012-2014&2010-2012&2012-2014
\\ \cmidrule(lr){3-4} \cmidrule(lr){5-6} \cmidrule(lr){7-8}
\multirow{2}{*}{\rotatebox[origin=c]{90}{\footnotesize Train}} & \multicolumn{1}{|c}{2010-2012} & 40.90 & 35.57 & \textbf{43.11} $^\blacktriangleup$ & 34.12 & 41.83 & \textbf{40.02}$^\blacktriangleup$
\\
 & \multicolumn{1}{|c}{2012-2014} & 30.51 & 44.96 & 30.38 & 47.18 & \textbf{39.11}$^\blacktriangleup$ & \textbf{47.28}
\\\bottomrule 
\end{tabularx}
\vspace{-10pt}
\end{table*}

To evaluate the transferability of the models, besides the debates from the last period of Dutch parliament, we have used debates from October 2010 to March 2012 where VVD and CDA were pro-government parties and others were oppositions.  We use $SVM$ as the base classifier to predict party that each member belongs to, give the speaches of the members. We have done classification using the $SVM$ itself as well as using $SVM$ by considering probabilities of terms in \actdsm as the weights of features in order to evaluate the effectiveness of \actdsm as the separable representation of data. This way, we make use of \actdsm like a feature selection approach that filters out features that are not essential in accordance to the hierarchical position of entities and make the data representation more robust by taking out non-stable terms. 
We have also tried $SVM$ along with other feature selection methods~\citep{Forman:2003,brank:2002} as the baselines, here we report the results of using Information Gain (IG) as the best feature selection method in our task in the parliament dataset.
We have employed conventional 5-fold cross validation for training and testing and to maintain comparability, we have used the same split for folding in all the experiments.
Tables~\ref{tbl:party} shows the performance of $SVM$, $SVM_{IG}$, and $SVM_{\actdsm}$ on party classification over two periods in terms of macro-average accuracy.
Comparing the results, it can be seen that $SVM_{\actdsm}$ improves the performance of classification over $SVM$ in all the experiments.  

Although $SVM_{IG}$ performs very well in terms of accuracy in within period experiments, it fails to learn a transferable model in cross period experiments and even it performs a little bit worse than the $SVM$ itself. We looked into the confusion matrices of cross-period experiments and observed that most of the errors in both $SVM$ and $SVM_{IG}$ are because of misclassified members of CDA to PvdA and vice versa. These are the two parties that their statuses have been changed in these periods. 

We investigate models of these two parties to understand how separation in the feature representation affects the performance of cross period classification. To do so, for each of these two classes, in each period, we extract three probability distributions on terms indicating their importance based on different weighting approaches: 1) Term Frequency (used as feature weights in $SVM$), 2) Information Gain (used as feature weights in $SVM_{IG}$), and 3) probability of terms in \actdsm (used as feature weights in $SVM_{\actdsm}$). 
Then, as a metric to measure separability of features, we use the Jensen-Shannon divergence to calculate diversity of probability distributions in three cases: 1) Different Parties in the Same Period, 2) Same Party in Different Periods 3) Different Parties in Different Periods. 
To avoid the effect of the number of features on the value of divergence, we take the top 500 high scored terms of each of weighting methods as the fixed length representatives of them. Figure~\ref{fig:Chart} shows the average diversity of distributions in each of the three cases for each of the three weighting methods. 

As expected, the diversity of features for different parties in a same period is high for all the methods and $IG$ provides the more separable representation in this case, which results in its high accuracies in within period experiments. However, when we calculate the diversity of features for a same party in different periods, feature representations are different in both $TF$ and $IG$, which causes false negative errors in the classification of these two parties. An interesting observation is in the case of having different parties in different periods, while we have two different parties their feature representations are similar in both $TF$ and $IG$, which leads to false positive errors in the classification. 

Considering these observations together reveals that $SVM$ and $SVM_{IG}$ learn models on the basis of features that are indicators of issues related to the status of parties, since they are the most discriminating terms considering one period and in within period experiments, the performance of $SVM_{IG}$ and $SVM$ is indebted to the separability of parties based on their statuses. Hence, after changing the status in the cross period experiments the trained model of the previous period generated by $SVM$ and $SVM_{IG}$ fails to predict the accurate party.  In the same way, the status classifier is affected by different parties forming a government in different periods, leading to lower accuracies.   

This is exactly the point which the strengths of \actdsm kicks in. In fact, two\:-\:dimensional separability in the feature representation, enables $SVM$ to tackle the problem of having non-stable features in the model when the status of a party changes over time. In other words, eliminating the effect of the status layer in the party model, which is the result of the horizontal separation, ensures that the party model captures the terms related to the party ideology, not its status. Thereby, not only $SVM_{\actdsm}$ learns an effective model with acceptable accuracy in within period experiments, but also its learned  models remain valid when the statuses of parties change. 
\pgfplotstableread{
0    0.4801  0.8616  0.7492  
1    0.5667  0.5941  0.2422
2    0.3422  0.3222  0.6822
}\dataset

\definecolor{b}{HTML}{4981CE}
\definecolor{g}{HTML}{859C27}
\definecolor{r}{HTML}{B22222}
\definecolor{o}{HTML}{FF6600}

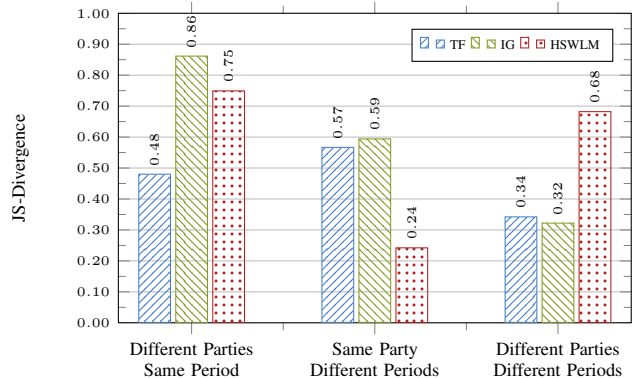
\begin{figure}[!t]
\centering
\begin{tikzpicture}
\pgfkeys{
    /pgf/number format/precision=2, 
    /pgf/number format/fixed zerofill=true,
    /pgf/number format/fixed
}
\begin{axis}[
    width= 8.4cm, 
    height=5.7cm, 
    enlarge y limits=0.0,
    enlarge x limits=0.2,
    ymajorgrids,
    minor tick num=1,
    ybar,
    bar width= 12pt,
    xtick=data,
    xticklabel style = {font=\fontsize{7}{8}\selectfont, align=center, text width=1.8cm},
    xticklabels = {Different Parties Same Period, Same Party Different Periods, Different Parties  Different Periods},
    ymin=0.0, 
    ymax=1.0,
    ytick = {0.00,0.10,0.20,0.30,0.40,0.50,0.60,0.70,0.80,0.90, 1.0},
    label style = {font=\fontsize{7}{8}\selectfont, yshift=0.5ex},
    anchor=north,
    ylabel={JS-Divergence},
    legend style={at={(0.965,0.95),font=\fontsize{5}{6}\selectfont},
    legend columns=-1
    },
    nodes near coords,
    every node near coord/.append style={font=\fontsize{5}{6}\selectfont, rotate=90, anchor=west},
    tick label style={font=\tiny},
    ]

\addplot[fill=b, draw=b, pattern color = b, pattern = north east lines] table[x index=0,y index=1] \dataset; 

\addplot[fill=g, draw=g, pattern color = g, pattern = north west lines] table[x index=0,y index=2] \dataset;

\addplot[fill=r, draw=r, pattern color = r, pattern = dots] table[x index=0,y index=3] \dataset; 

\legend{TF,IG,\actdsm}

\end{axis}
\end{tikzpicture}
\vspace{-15pt}
\caption{
Average diversity of the representation of features of CDA and PvdA in different situations\label{fig:Chart}}
 \vspace{-25pt}
 \end{figure}

We furthermore looked into the size of estimated {\actdsm}s by the number of terms with non-zero probability and, on average, the size of the models are about 100 times smaller than the number of features selected by $IG$ in the corresponding models. 
So, although \actdsm takes considerable risk of loosing accuracy in within period experiments by aggressively pruning the overlapping terms, it provides small and precise models that are not only effective over time, but also efficient when the size of data is large.

\smallskip
In summary, we demonstrated that \actdsm indeed exhibit the two\:-\:dimensional separation property for the parliamentary data, as predicted by our theoretical analysis in the earlier sections. In addition, we empirically validated the transferability of models using \actdsm which is the result of its two\:-\:dimensional separability.

\vspace{-5pt}
\shrink 
\section{Conclusions}
\label{sec:con}
\sshrink
In this paper, we investigated the separation property in hierarchical data focusing on hierarchical text classification.

Our first research question was: ``\textsl{\qone}''
We demonstrated that based on the ranking and classification principles, the \emph{separation property} in the data representation is a desirable foundational property which leads to separability of scores and consequently improves the accuracy of classifiers' decisions.  We stated this as the ``\ssp'' for optimizing expected effectiveness of classifiers.

Our second research question was: ``\textsl{\qtwo}''
We showed that in order to have horizontally and vertically separable models, they should capture all, and only, the essential terms of the entities taking their position in the hierarchy into account. Based on this, we introduced \TDSMs for estimating separable models for hierarchical entities. We investigated \actdsm and demonstrated that it offers separable distributions over terms for different entities both in case of being in the same layer or in different layers.

Our third research question was: ``\textsl{\qthree}''
We evaluated the performance of classification over time using separable representation of data and showed that separability makes the model more robust and transferable over time by filtering out non-essential non-stable terms.

The models we proposed in this paper 
are IR models which are applicable to a range of information access tasks~\citep{Dehghani:2016:SIGIR,Dehghani:2016:CIKM1,Dehghani:2016:CLEF,Dehghani:2016:CHIIR}, not just hierarchical classification, as many complex ranking models combine different layers of information.
There are number of extensions we are working on in future work.
First, in this research we focused on text\:-\:dominant environment and considered all terms in the text as features. However, this can be done considering terms with a specific part of speech, 
or even on non\:-\:textual models with completely different types of features.
Second, it would be beneficial to construct mixture models for terminal entities using \actdsm in a way that constructed mixture models are capable of reflecting local interactions of terminal entities in different layers.

\sshrink
\sshrink
\mypar{Acknowledgments}
%
%
\small{
This research is funded in part by Netherlands Organization for Scientific Research through the \textsl{Exploratory Political Search} project (ExPoSe, NWO CI \# 314.99.108), and by the Digging into Data Challenge through the \textsl{Digging Into Linked Parliamentary Data} project (DiLiPaD, NWO Digging into Data \# 600.006.014).}

\vspace{-5pt}
%
\renewcommand{\bibsection}{\shrink\section*{References}\sshrink}
\bibliographystyle{abbrvnat}
\raggedright
\renewcommand{\bibfont}{\small}
\setlength{\bibhang}{0pt} 
\setlength{\bibsep}{0pt} 
\bibliography{ref}  
%
%
\balancecolumns
\end{document}